\documentclass[aps,prb,reprint,linenumbers,amssymb,byrevtex]{revtex4-1}

\usepackage[T1]{fontenc}
\usepackage{graphicx}
\usepackage{bm}

\begin{document}

\title{'Kondo state' and Kondo resonance in a two-dimensional electron gas}

\author{Stefan M\"{u}llegger}
\email{stefan.muellegger@jku.at}

\author{Mohammad Rashidi}
\author{Michael Fattinger}
\author{Reinhold Koch}
\affiliation{Institute of Semiconductor and Solid State Physics, Johannes Kepler University, Linz, Austria.} 

\begin{abstract}
The delicate balance of spin-screening and spin-aligning interactions determines many of the peculiar properties of dilute magnetic systems. 
We study a surface-supported all-organic multi-impurity Kondo spin system at the atomic scale by low-temperature scanning tunnelling microscopy and -spectroscopy. 
The model system consists of spin-1/2 radicals that are aligned in one-dimensional chains and interact \textit{via} a ferromagnetic RKKY interaction mediated by the 2DEG of the supporting substrate. 
Due to the RKKY-induced enhanced depopulation of one spin-subband in the 2DEG, we finally succeeded to detect the so far unobserved 'Kondo state' as opposed to the well-established Kondo resonance. 
Its cloud of screening electrons, that are virtually bound to the radicals below the Kondo temperature, represents the extended exchange hole of the ferromagnetically polarized spin chain imaged here in real space. 
\end{abstract}

\maketitle

The interaction of localized electron spins of magnetic impurities with delocalized conduction electrons provides a key handle for controlling the spin polarization in diluted magnetic systems essential for future spintronics \cite{Wolf2001,Milestones2008} and molecular electronics \cite{Scott2010} applications. 
Many of their unique properties are determined by the competition between screening of local spins and magnetic interactions of neighbouring spins. 
Below a characteristic temperature the impurity spin may be screened by a collective of conduction electrons with opposite spin -- a phenomenon known as the Kondo effect \cite{Kondo1964}. 
The screening electrons form a virtual bound singlet many-body quantum state with the singly occupied impurity \cite{Hewson1993} denoted herein as the 'Kondo state' \cite{Scott2010}. 
At higher impurity concentrations, \textit{i.\,e.} at sufficiently small separations, their individual Kondo states may overlap, causing an additional magnetic RKKY\cite{Kittel1968}-type interaction of individual impurity spins \cite{Simon2005} that competes with Kondo screening. 

\begin{figure}
\includegraphics[width=7.5cm]{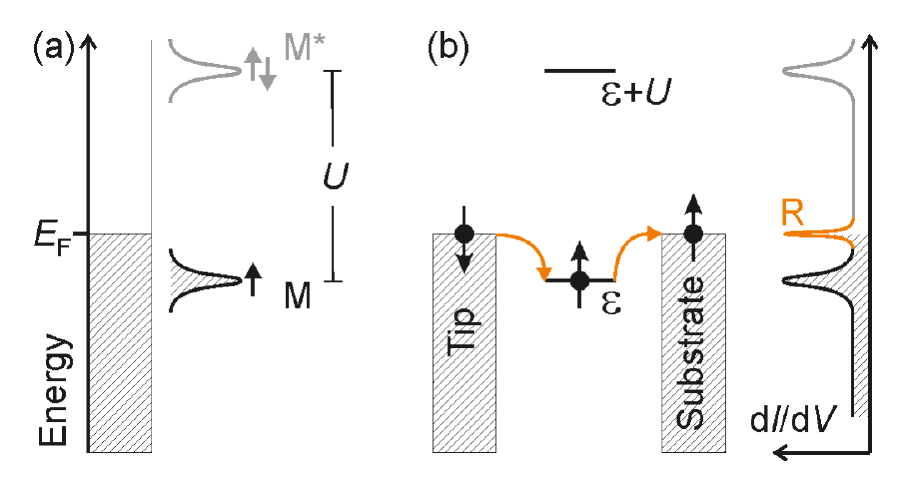}
\caption{\label{fig:wingreen} \textbf{Kondo resonance.} (a) Energy level configuration of the Anderson single-impurity model \cite{Anderson1961}. In this model the energy $\epsilon$ of the magnetic impurity level (M) is assumed to lie sufficiently far below $E_\mathrm{F}$ so that its occupation probability $p$ is close to one, thus avoiding mixed valency ($p<1$). 
Double occupation of the impurity level costs the Coulomb charging energy $U$ and raises this state to M$^*$ well above $E_\mathrm{F}$, thus making addition of a second electron energetically unfavourable.
In the bulk of metals containing dilute magnetic impurities the Kondo effect gives rise to a higher density of states \cite{Hewson1993} for conduction electrons with opposite spin localized at the impurity sites. 
As a result the scattering cross-section for the conduction electrons is increased, which explains the resistivity increase of bulk Kondo systems below the Kondo temperature $T_\mathrm{K}$ \cite{Kondo1964}.
(b) In the case of tunnelling transport the low-lying impurity state opens up a new transport channel close to $E_\mathrm{F}$ \cite{Ujsaghy2000,Wingreen2001,Plihal2001,Scott2010}. 
Whenever an electron tunnels off the impurity, leaving behind an empty virtual intermediate state, it is replaced almost instantaneously by a conduction electron of opposite spin as demanded by the high occupation probability close to one. 
The extra transport channel increases \cite{note1} the transmission (conductance $\mathrm{d}I/\mathrm{d}V$) close to $E_\mathrm{F}$ and is commonly denoted Kondo resonance (R) in the literature \cite{Wingreen2001,Scott2010}.}
\end{figure}

Recently, the study of single impurity spins on atomic scale has become a reality by means of electron transport through lateral quantum-dots \cite{Goldhaber1998,Cronenwett1998,Jeong2001} formed in a two-dimensional electron gas (2DEG) as well as surface-supported $d$-metal impurities \cite{Li1998b,Madhaven1998,Ternes2009} or metalorganic molecules \cite{Iancu2006a,Gao2007} inside the tunnel junction of a scanning tunnelling microscope (STM). 
Studies of single Kondo impurities typically report a resonance observed in the electron transmission spectrum near the Fermi energy, $E_\mathrm{F}$, \cite{Wingreen2001} which is commonly rationalized by the Anderson single-impurity model \cite{Anderson1961} of localized magnetic states in metals [see detailed description of Fig.~\ref{fig:wingreen}]. 
Already in earlier work \cite{Manoharan2000} the resonance signal was utilized for obtaining spatial information on the 'Kondo screening cloud' \cite{Affleck2001,Affleck2008}, \textit{i.\,e.} the spatial region captured by the screening electrons surrounding the impurity. 
While the Kondo resonance involves only conduction electrons (near $E_\mathrm{F}$), the virtual bound screening state (=Kondo state) is formed by all electrons screening the localized spin, i.e. the conduction electrons near $E_\mathrm{F}$ as well as deeper lying band electrons. 
A direct detection of the complete screening cloud (exchange hole) representing the Kondo state itself, has remained elusive until now. 

Here we report on a low-temperature STM study of surface-supported dimers and chains of the organic $sp$ radical BDPA ($\alpha$,$\gamma$-bisdiphenylene-$\beta$-phenylallyl, C$_{33}$H$_{22}$) [Fig.~\ref{fig:kondo}a], which is a well-established spin-1/2 standard in electron spin resonance spectroscopy \cite{Duffy1972}. 
We finally succeeded to tunnel out of the Kondo state itself and thus were able to detect and directly visualize the exchange hole \cite{Schumann2005,Schumann2010} generated by a linear chain of molecular Kondo impurities in a 2DEG. 
In the present work, experimental detection of the Kondo state is enabled by a remarkably strong change in conductance facilitated by the exceptional combination of properties of our Kondo system -- as opposed to atomic $d$-metal Kondo systems in the literature: 
(i) Weak physisorption of the radicals on the substrate preserves their magnetic state. 
(ii) The reconstructed (111) surface of Au provides the crucial combination of surface anisotropy (herringbone reconstruction) together with a Shockley-type surface state near $E_\mathrm{F}$. 
The former guides the self-assembly of structurally well-ordered BDPA chains and the latter acts as a 2DEG capable of communicating the Kondo effect over large distances of more than a nanometre across the sample surface (typical for coinage metal (111) surfaces \cite{Manoharan2000,Henzl2007,Li2009}). 
Thus individual Kondo states of neighbouring BDPAs are allowed to overlap.
(iii) Since the singly occupied highest molecular orbital (SOMO) of BDPA is symmetry-compatible with the $sp$ wave functions of the 2DEG, constructive interaction of neighbouring spins is facilitated. 
(iv) The radical-radical separation in BDPA chains on Au(111) is appropriate to give rise to an additional ferromagnetic \cite{Simon2005,Wahl2007} RKKY-type interaction of individual BDPA spins, preferentially mediated by the 2DEG, as well. 
(v) The resulting constructive superposition of individual Kondo states leads to an effective depopulation of one spin-subband and thus an enhancement of the conductance change enabling detection of the joint Kondo state by the STM tip. 

\subsection*{BDPA self-assembly on Au(111)}

\begin{figure}
\includegraphics[width=8.6cm]{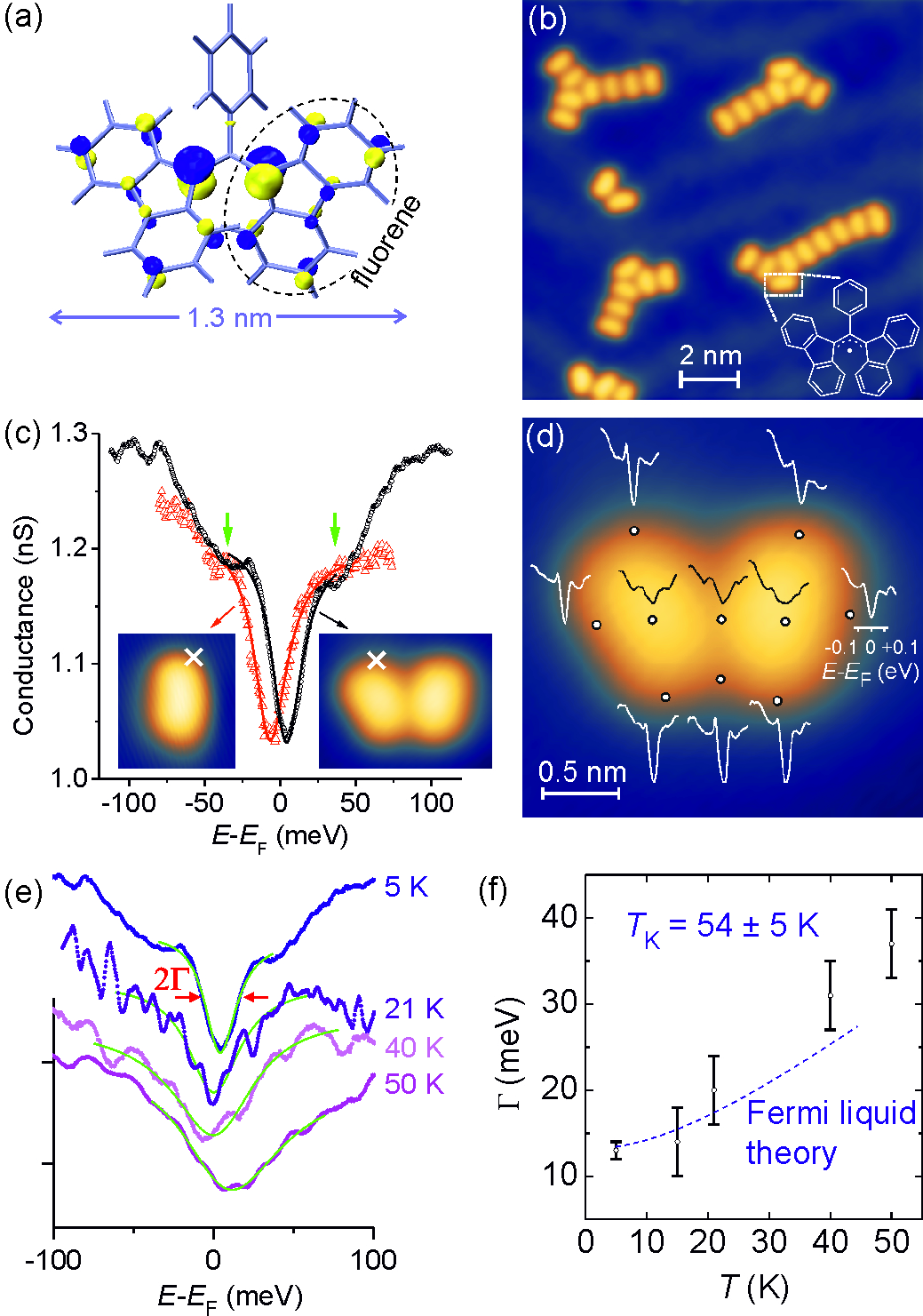}
\caption{\label{fig:kondo} \textbf{BDPA on Au(111).} (a) Chemical structure and DFT-calculated SOMO spin density of BDPA; the two 'hot spots' are centred around the meso-carbons. (b) STM topograph of self-assembled BDPA clusters on the Au(111) surface (+20~mV, 200~pA). (c) $\mathrm{d}I/\mathrm{d}V$ spectrum of a single BDPA radical (red) and a dimer (black); the zero-bias conductance of the pristine Au(111) background at the Fermi level was calibrated to about $1.3$~nS; solid lines are numerical Fano fits of Kondo antiresonance (see also Tab.~\ref{tab:fano}); green arrows mark vibrational satellites; insets show STM topographs of single BDPA and dimer; crosses mark the STM tip position for recording $\mathrm{d}I/\mathrm{d}V$ spectrum. (d) $\mathrm{d}I/\mathrm{d}V$ Kondo spectra over different positions ($\circ$) of a BDPA dimer. (e) $\mathrm{d}I/\mathrm{d}V$ Kondo spectra of a BDPA dimer at different temperatures. (f) Temperature-dependent broadening; dashed line represents best-fit based on formalism of Nagaoka \textit{et al.} \cite{Nagaoka2002}.}
\end{figure}

The STM topograph of Fig.~\ref{fig:kondo}b reveals the formation of benzene-free BDPA nanostructures upon deposition onto Au(111) at 300~K. 
Individual BDPA radicals are imaged as bean-shaped protrusions [see also inset of Fig.~\ref{fig:kondo}c]. 
Pronounced clustering dominates already at submonolayer coverages, and dimers of BDPA constitute the smallest clusters. 
Larger BDPA nanostructures exhibit characteristic star-like shapes consisting of three branches of one-dimensional BDPA chains up to several nanometres long (depending on coverage). 
Single BDPAs are rarely observed on the surface due to an increased surface mobility even at 5~K, which strongly constrains the energy range accessible by STS compared to dimers and chains. 
As determined from STM topographs, BDPAs in chains pack closer (0.73~nm) than in dimers (0.8--0.9~nm). 
In the bulk phase BDPA is a one-dimensional paramagnet above 1.7~K \cite{Duffy1972} forming similar linear chains with a significantly larger radical-radical spacing of 1.4~nm along the $\vec{b}$ crystallographic direction. 

\subsection*{Kondo resonance}

\begin{table}
\caption{\label{tab:fano} Fano fit parameters of the BDPA monomer (average of four), dimer (average of three), and chain (average of three) adsorbed on Au(111) at 5~K; asymmetry parameter $q$, resonance position $\Delta E$ and half-width $\Gamma$.}
\begin{tabular}{lccc}
 &  $q$ ($\pm0.1$~) & $\Delta E$~($\pm0.7$~meV) & $\Gamma$~($\pm1$~meV) \\
\hline
monomer & $-0.15$ & $-5.9$ & $11$ \\
dimer & $-0.23$ & $+2.3$ & $13$ \\
chain & $-0.15$ & $+7.0$ & $19$ \\
\hline
\end{tabular}
\end{table}

The unpaired electron of BDPA resides in the SOMO that derives from $sp$ hybrid orbitals and is delocalized over the fluorene units of the molecular backbone [Fig.~\ref{fig:kondo}a]. 
When the STM tip is located over BDPA adsorbed on Au(111), typical $\mathrm{d}I/\mathrm{d}V$ spectra exhibit a strong conductance minimum (dip) close to $E_\mathrm{F}$ [Fig.~\ref{fig:kondo}c]. 
The maximum amplitude is observed for tip positions near the rim of BDPA, while over intramolecular positions the dip broadens and decreases [Fig.~\ref{fig:kondo}d]. 
The dip amplitude amounts to more than 20\% of the zero-bias conductance of the pristine substrate for our best STM tips. 
We observe almost identical spectra for BDPA radicals adsorbed over different fcc and elbow positions of the reconstructed Au(111) surface, suggesting that the adsorption site relative to the substrate atomic lattice is negligible.
A detailed analysis of our STS results confirms the characteristic properties of the transmission Kondo effect \cite{Li1998b,Madhaven1998} for BDPA/Au(111): 
(i) The conductance minimum lies within a few milli-electronvolts around $E_\mathrm{F}$. 
(ii) The resonance width, $2\Gamma$, increases with temperature and the amplitude decreases simultaneously [Figs.~\ref{fig:kondo}e,f]. 
(iii) The spectral shape of the dip \cite{note1}  is well described by the Fano function \cite{Fano1961,Ujsaghy2000,Figgins2010} [Fig.~\ref{fig:kondo}c]; the numerical values of the fit parameters are listed in Tab.~\ref{tab:fano}. 
(iv) The resonance is accompanied by two satellite features symmetrically offset around the Kondo dip [marked by green arrows in Fig.~\ref{fig:kondo}c] indicating molecular vibrations excited by inelastic tunnelling processes. 
This is expected for molecule-based Kondo systems where the impurity orbital (SOMO) is spatially extended and thus sensitive to geometric changes of the molecular backbone \cite{Torrente2008}. 
The respective vibrational energy of $\approx$\,35--55~meV (300--450~cm$^{-1}$) points to one or more collective vibrational modes of the BDPA backbone rather than plain C-H vibrations, which are typically observed at higher energies as suggested by our density functional theory (DFT) calculations. 

The lower surface mobility of BDPA dimers compared to single radicals facilitates recording of the Kondo resonance width up to temperatures of 50~K [Fig.~\ref{fig:kondo}f].  
The respective broadening is analysed in detail based on a Fermi liquid description put forward by Nagaoka \textit{et al.} \cite{Nagaoka2002}. 
The least-square fit based on this formalism, $\Gamma = 2 \sqrt{(\pi k_\mathrm{B} T)^2+2(k_\mathrm{B} T_\mathrm{K})^2}$, is shown as dashed curve in Fig.~\ref{fig:kondo}f.
The obtained nominal value of $T_\mathrm{K}=54\pm5$~K is consistent with the resonance width of $2k_\mathrm{B} T_\mathrm{K}$ predicted by Fermi liquid theory. 
[For further details on the Kondo signal of BDPA/Au(111) see Supplementary Information S1.]

The existence of a Kondo signal corroborates that BDPA preserves the magnetic ground state after adsorption on Au(111) at the single radical level. 
Our electron spin resonance (ESR) experiments on samples with BDPA monolayer coverage on Au(111)/mica substrates yield $g=1.96$ at $7$~K. 
This value is in good agreement with that of the single-crystal bulk phase ($2.0026$) \cite{Hamilton1963} as well as that observed in ultrathin BDPA films drop-cast on Au(111) substrates (2.005) \cite{Messina2007} at room temperature. 
Both, Kondo behaviour and ESR signal, evidence that a possible charge transfer between BDPA and the Au substrate is very small. 

\begin{figure}
\includegraphics[width=8.6cm]{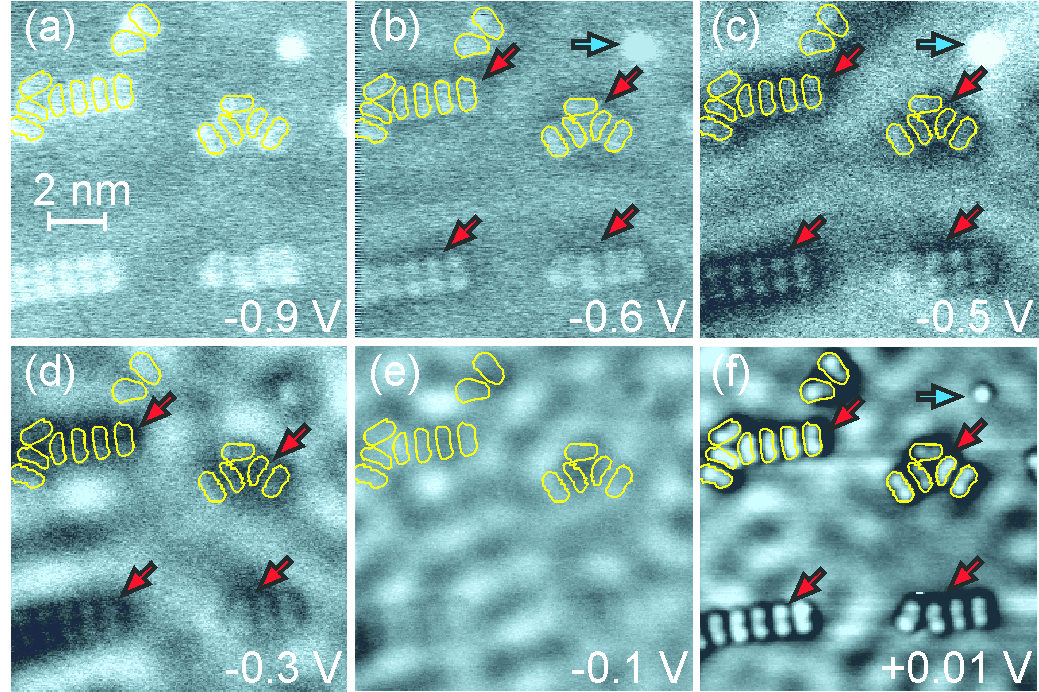}
\caption{\label{fig:maps} \textbf{Spectroscopic $\mathrm{d}I/\mathrm{d}V$ maps at different bias voltages.} Overlayed yellow contour lines mark areas covered by BDPA molecules as extracted from STM topographs; red arrows indicate reduced-conductance envelopes, blue arrows a surface defect.}
\end{figure}

\subsection*{Experimental observation of the Kondo state}

The spectroscopic images of Fig.~\ref{fig:maps} reveal intriguing details of the Kondo screening of BDPA clusters on Au(111).
Doubly occupied pure MO states [see Supplementary Information S2] are imaged as bright protrusions confined to the area covered by the BDPA molecules [Fig.~\ref{fig:maps}a]. 
For sample bias ranging from about $-0.6$ to $-0.3$~V we observe areas of reduced conductance (dark) enveloping the BDPA clusters as marked by red arrows in Figs.~\ref{fig:maps}c,d. 
These areas clearly extend more than 1~nm into regions of the pristine Au(111) surface outside of and in between neighbouring BDPAs -- in clear contrast to the pure MO states of Fig.~\ref{fig:maps}a. 
We denote them as reduced-conductance envelopes (RCEs), since $\mathrm{d}I/\mathrm{d}V$ is reduced by about 50\% relative to zero-bias conductance of the pristine Au(111) surface. 
The RCEs start to appear slightly below the onset of the Au(111) surface-state band close to $-0.5$~V [Figs.~\ref{fig:maps}b,c].
They persist over an energy range of about 200~meV [Fig.~\ref{fig:maps}d] and fade off when further approaching zero bias.
Above $-0.3$~V the pristine standing-wave pattern of the surface state electrons is imaged [Fig.~\ref{fig:maps}e].
It originates from scattering at the herringbone reconstruction pattern \cite{Burgi2002} of the Au(111) surface and the BDPA clusters. 
In the standing wave pattern of Fig.~\ref{fig:maps}e we find no evidence for an enhanced electrostatic scattering potential localized at the BDPA radicals (which are hardly visible). 
This evidences \cite{Olsson2007} that the adsorbed radicals have (almost) zero total charge -- in agreement with our ESR results. 
Even with the STM tip above BDPA the characteristic step-like conductance signal of the surface state survives almost unaffected [see Figure S1 of the Supplementary Information]. 
This indicates that hybridization of BDPA orbital(s) with the surface state is only weak and, in particular, rules out that the observed RCEs originate from a possible localization \cite{Limot2005} of the surface state. 
The possibility that local interference minima of the elastically scattered surface-state electron waves would cause the observed RCEs is certainly ruled out \cite{Crommie2000,Schneider2005}, as well:
(i) The RCEs are absent in the surface-state standing-wave pattern of Fig.~\ref{fig:maps}e. 
(ii) The RCEs lack any oscillating behaviour expected for scattered surface-state waves \cite{Gross2004}. 
(iii) The RCEs are detected only at the BDPA units, but not at surface impurities [marked by a blue arrow in Figs.~\ref{fig:maps}b,c,f]. 
Possible life-time effects are known to flatten the surface-state band edge typically in the order of 10~meV \cite{Schneider2000} and thus are one order of magnitude too small to explain the observed RCEs, as well. 

Closely around zero bias the RCEs reappear again [Fig.~\ref{fig:maps}f] and represent the conductance map of the Kondo resonance measured point-by-point in Figs.~\ref{fig:kondo}c,d. 
The RCEs of Fig.~\ref{fig:maps}f thus reflect the lateral spatial distribution of the Kondo resonance signal of BDPA clusters on Au(111). 
The sample area at which the Kondo resonance is observed by STM has been commonly attributed to the spatial extent of the Kondo screening cloud \cite{Manoharan2000,Pruser2011}.
However, the interpretation of its size remains a controversial matter, because it is much smaller than the Kondo screening length predicted by Fermi liquid theory [see Supplementary Information S3]. 
The Kondo dip of BDPA/Au(111) is detected more than 1.5~nm laterally away from BDPA and decays according to $1/r$ [see Supplementary Information~S4]. 
This suggests that the Kondo screening of BDPA/Au(111) strongly involves delocalized electronic states \cite{Plihal2001,Knorr2002} -- most possibly the surface-state band of Au(111) that acts as a 2DEG.

\subsection*{Discussion: Kondo state vs. Kondo resonance}

\begin{figure}
\includegraphics[width=8cm]{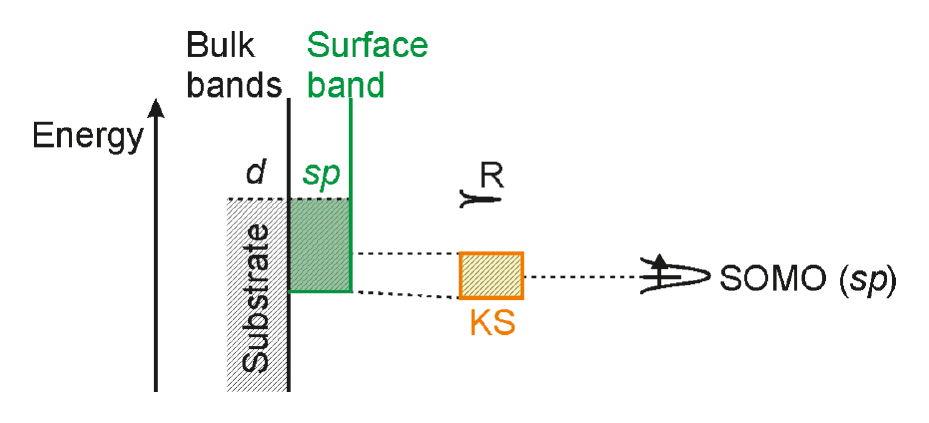}
\caption{\label{fig:model} \textbf{Schematics summarizing the $\mathrm{d}I/\mathrm{d}V$ results of BDPA/Au(111).} In addition to the Kondo resonance (R) close to $E_\mathrm{F}$ the Kondo state (KS) is detected experimentally via contributions from deeper lying surface-state band electrons. The Kondo state originates from resonant spin exchange processes between the magnetic impurity (SOMO of BDPA) and the surface-state band of Au(111).} 
\end{figure}

The additional RCEs of Figs.~\ref{fig:maps}c,d observed at energies well below $E_\mathrm{F}$ come -- at first sight -- as a surprise. 
Since the most obvious possible alternatives are ruled out above, there is strong evidence that the additional RCEs originate from the Kondo state (KS) itself, i.e. the many-body virtual bound state including the spin-screening contributions from deeper lying band electrons as well as conduction electrons near $E_\mathrm{F}$. 
This is illustrated in Fig.~\ref{fig:model}:  
(i) Their energy onset coincides remarkably well with the band edge of the Au(111) surface-state.
The latter is strongly involved in the Kondo screening of BDPA/Au(111), as indicated by our results.  
(ii) The geometric size and shape are qualitatively similar to the RCEs of the Kondo resonance, but the signal amplitude is much stronger and they extend even further into areas of the pristine Au(111) surface [compare Figs.~\ref{fig:maps}d and f]. 
(iii) The Kondo state involves an interaction of the surface-state and the SOMO, suggesting a similar energy range for both. 
This is illustrated in Fig.~\ref{fig:model} and corroborated by point spectroscopy [see Supplementary Information S2]. 

The observed reduction of the tunnelling conductance is rationalized as follows:
The density-of-states of the 2DEG consists of two overlayed identical step-functions \cite{Datta2005} with different Fermi wave vectors, Rashba-split in $k$-space \cite{Reinert2003} but having the same energy onset. 
Our results indicate that Kondo screening depletes one of the spin subbands in the surface area around BDPA (\textit{e.\,g.} the up-spin subband if BDPA is up-spin). 
The decreased occupation of one subband (exchange hole) reduces the number of available electrons for tunnelling out of the surface state. 
The spin screening is effective not only for conduction electrons near $E_\mathrm{F}$, but for deeper lying band electrons, as well. 
This equals an effective loss of a tunnelling transmission channel and thus a reduction of the tunnelling conductance -- as evidenced by our experiments. 

\begin{figure}
\includegraphics[width=8.6cm]{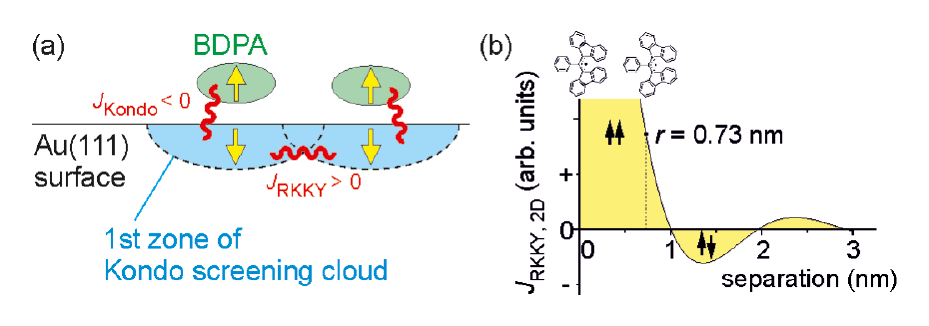}
\caption{\label{fig:rkky} \textbf{Competition of RKKY and Kondo interaction.} (a) Schematics illustrating the overlap of the Kondo clouds of two neighbouring BDPA radicals adsorbed on the Au(111) surface. (b) Calculated separation-dependent oscillatory behaviour of the two-dimensional RKKY coupling constant $J_\mathrm{RKKY,\,2D}$ for $k_\mathrm{F}^{-1}=0.614$~nm.}
\end{figure}

The $\mathrm{d}I/\mathrm{d}V$ images of Figs.~\ref{fig:maps}c,d corroborate that the Kondo state (\textit{i.\,e.} the screening cloud) of individual BDPA radicals on Au(111) are so large that they readily overlap with those of their next neighbours. 
In the case of two or more neighbouring Kondo impurities an additional RKKY-type indirect-exchange interaction of their spins evolves, according to predictions based on Fermi-liquid theory \cite{Simon2005}. 
The RKKY interaction of neighbouring Kondo impurities is either ferromagnetic ($J_\mathrm{RKKY}>0$) or antiferromagnetic \cite{Iancu2006a,Wahl2007,Tsukahara2011,Bork2011} (<0) and competes with Kondo screening ($J_\mathrm{Kondo}<0$) as illustrated in Fig.~\ref{fig:rkky}a. 
Similar to the Kondo screening, the RKKY interaction between BDPA radicals in chains on Au(111) is predominantly mediated by surface-state electrons. 
The surface band of Au(111) starts 500~meV below $E_\mathrm{F}$, and has a Fermi wave number of $k_\mathrm{F}^{-1}=0.617$ and $0.540$~nm including the Rashba splitting \cite{Reinert2003}. 
A possible coupling mediated by bulk conduction electrons decays at much shorter distances due to the smaller $k_\mathrm{F}^{-1}=0.08$~nm \cite{Reinert2003} and thus is negligible for intermolecular interactions in BDPA clusters. 
A calculation of the distance-dependent oscillation of the RKKY coupling constant for the two-dimensional case \cite{Beal1987} based on the momentum and effective mass ($m^*=0.25 m_\mathrm{0}$) \cite{Reinert2003} of the Au(111) surface state yields a ferromagnetic RKKY interaction for separations smaller than 1~nm.
Thus, for next-neighbour BDPA radicals in chains, which are separated by 0.73~nm, a parallel spin alignment is favoured as illustrated in Fig.~\ref{fig:rkky}b. 
Ferromagnetic RKKY facilitates a constructive superposition of individual Kondo states, which enhances the depopulation of one subband and, consequently, further lowers the conductance for tunnelling out of the surface-state. 
This enhancement effect seems to be essential for enabling experimental detection of the Kondo state. 
Since the RKKY interaction strength decays with $1/r^3$, the influence of the next-neighbour BDPA is dominant in BDPA chains. 
Thus BDPA chains on Au(111) are expected to be paramagnetic at 5~K in agreement with our results as well as a recent ESR study \cite{Messina2007}. 

Tunneling out of the Kondo state is further supported by the fact that the molecules are weakly imaged by STS between $-0.7$ and $-0.5$~V but not at $-0.1$~V.

In dimers the radical-radical separation of 0.8--0.9~nm is larger and close to the transition between ferro- and antiferromagnetic alignment [Fig.~\ref{fig:rkky}]. 
Consequently, the RKKY-based enhancement effect for the Kondo state signal of BDPA dimers is decreased -- consistent with our results [Fig.~\ref{fig:maps}c,d].
In the case of the mobile monomers, we did not succeed to detect the Kondo state, although the Kondo resonance close to $E_\mathrm{F}$ is clearly observed [Fig.~\ref{fig:maps}f]. 
This is consistent with the absence of a RKKY-based enhancement effect in monomers.
The interaction between neighbouring radicals in the chains is observed to affect the width of the spectral Kondo resonance. 
Table~\ref{tab:fano} lists the values of the resonance width, $\Gamma$, for the BDPA monomer, dimer and chain as determined from our experiments.

\subsection{Conclusions}

We succeeded to detect and image the joint Kondo state (as opposed to the Kondo resonance) of an all-organic multi-impurity Kondo system at the atomic scale. 
Our Kondo system consists of spin-1/2 radicals that are aligned in one-dimensional chains and ferromagnetically coupled \textit{via} RKKY interaction mediated by the 2DEG of the supporting substrate. 
The related cloud of screening electrons with opposite spin and virtually bound to the radicals below the Kondo temperature represents the extended exchange hole of the ferromagnetically coupled spin chain. 
Our achieved mapping of the Kondo state utilizes the RKKY-enhanced depopulation of one spin subband in the 2DEG and opens a new way to study magnetic interactions of spin-carrying molecules in coupled spin clusters with impact on the development of molecular-based magnetic logic applications for future spin- and molecular electronics.

\subsection*{Methods}
BDPA recrystallized in benzene was thermally evaporated in ultrahigh vacuum (UHV) from a quartz crucible at 383~K after thorough degassing at 373~K.
The single-crystal Au(111) surface was prepared by repeated cycles of 0.5~keV Ar$^+$ bombardment and annealing at 720~K.
STM and STS experiments were carried out at 7~K employing electrochemically etched W tips. 
The $\mathrm{d}I/\mathrm{d}V$ signal was obtained from the first-harmonic current signal detected by lock-in technique (0.5-2~kHz; 5-20~mV sinusoidal peak-to-peak voltage; average of 10 single spectra). 
W tips were deoxidized by annealing in UHV; impurity- and tip effects were minimized by careful sample preparation and multiple tip formings between the $\mathrm{d}I/\mathrm{d}V$ experiments, resulting in Au-coated STM tips. 
Reliable tip performance was established by accurately reproducing the $\mathrm{d}I/\mathrm{d}V$ signature of the Au(111) surface state from literature \cite{Burgi2002}. 
$\mathrm{d}I/\mathrm{d}V$ spectra were recorded at constant-height conditions, the spectroscopic images ($\mathrm{d}I/\mathrm{d}V$ maps) simultaneously with constant-current STM topographic imaging.
Gas-phase density functional theory (DFT) single point energy calculations were performed with Gaussian~03 package \cite{Gaussian2003} using B3LYP hybrid functional \cite{Becke1993}, 6-31G(d) basis set and fixed molecular geometry extracted from BDPA bulk crystal structure \cite{Azuma1994}. 

\subsection*{References}

\subsection*{Acknowledgements}

We thank Wolfgang Jantsch at Johannes Kepler University for the EPR investigations and Andreas Ney for stimulating discussion. 
Financial support by the Austrian Science Fund (FWF project P20773) is acknowledged. 

\subsection*{Author contributions}

S.M., M.R. and M.F. carried out experiments. S.M. and M.R. analyzed the data. S.M. and R.K. wrote the manuscript, planned and supervised the project. All authors discussed the manuscript. 

\subsection*{Additional information}

The authors declare no cometing financial interests. Supplementary information accompanies this paper on www.nature.com/naturephysics. Reprints and permissions information is available online at http://www.nature.com/reprints. Correspondence and requests for materials should be addressed to S.M.

\end{document}


\title{Supplementary information: Kondo state and Kondo resonance in a two-dimensional electron gas}

\author{Stefan M\"{u}llegger}
\email{stefan.muellegger@jku.at}
\author{Mohammad Rashidi}
\author{Michael Fattinger}
\author{Reinhold Koch}
\affiliation{Institute of Semiconductor and Solid State Physics, Johannes Kepler University, Linz, Austria.} 

\maketitle

\subsection*{S1. Details of the Kondo resonance of BDPA/Au(111)}

The shape of the Kondo resonance is determined by the phase between different tunneling paths (into the Fermi sea or into the impurity level) and the particel-hole symmetry of the conduction band \cite{Figgins2010}. 
The observation of a dip (reduced conductance) rather than a peak is reflected by the numerical value of the asymmetry parameter $q$ in the fitted Fano-lineshape. 
$q$ depends on the ratio of tunneling amplitudes for different possible tunneling paths \cite{Ujsaghy2000,Plihal2001}. 
The obtained value of $q<<1$ (see Table~I) thus indicates predominant tunneling into the continuum of substrate states rather than the impurity state (which is here the SOMO) due to Coulomb repulsion. 

The energetic position, $\Delta E=E-E_\mathrm{F}$, of the Kondo dip is offset from the Fermi level.
Both magnitude and sign of $\Delta E$ are clearly affected by the presence of a second neighbouring BDPA molecule in dimers [see Fig.~2c and Table~I]. 
While in single BDPAs the Kondo dip lies below the Fermi level, it lies above it for dimers. 
A negative (positive) offset, $\Delta E$, indicates an occupied (empty) state. 
Fermi-liquid theory rationalizes the offset $\Delta E$ by the asymmetric level alignment of SOMO and SUMO relative to $E_\mathrm{F}$ (the so-called electron-hole asymmetry \cite{Hewson1993})   determined by the occupation number of the impurity level (SOMO). 
Without particle-hole symmetry the SOMO occupation number becomes $\neq 1$ and the Kondo peak gets displaced relative to the Fermi level. \cite{Hewson1993} 
When it is $<1$ the level is less than half-full and the Kondo resonance lies above $E_\mathrm{F}$ and below it for $>1$. 

The width of the Kondo dip increases by about 19\% in dimers compared to singles (see Table~I). 

\subsection*{S2. Point spectroscopy}

\begin{figure}
\includegraphics[width=8cm]{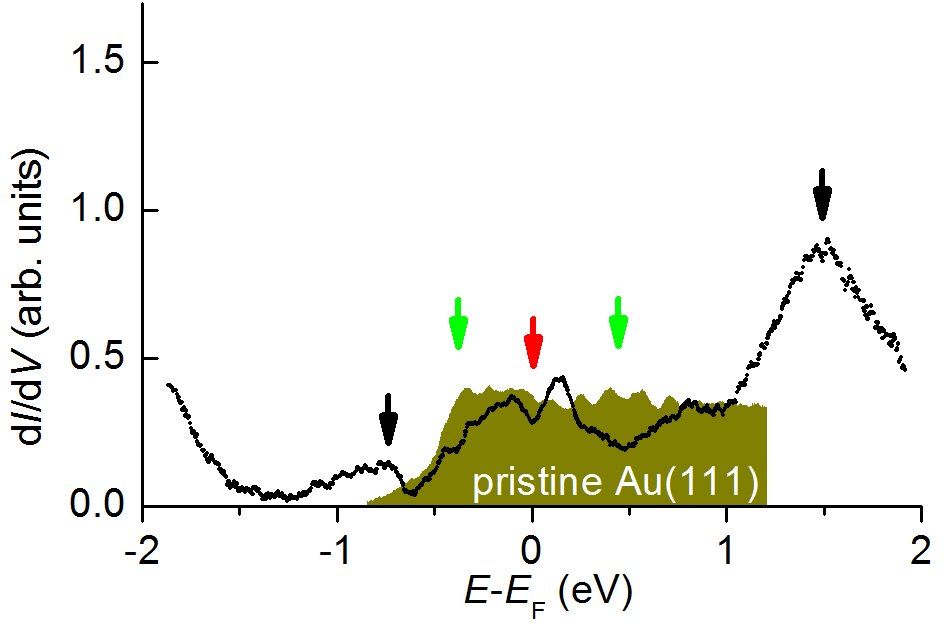}
\caption{\label{fig:spec} $\mathrm{d}I/\mathrm{d}V$ spectrum with STM tip over rim of BDPA dimer; dark yellow: pristine Au(111) surface. }
\end{figure}

The energy that determines whether Kondo physics will be visible is $k_\mathrm{B}T_\mathrm{K}\approx 5$~meV, is always smaller than the FWHM of the impurity level \cite{Hewson1993} (SOMO). 
To determine the energies and widths of SOMO and SUMO relative to the substrate Fermi level we recorded local $\mathrm{d}I/\mathrm{d}V$ spectra of BDPA dimers with the STM tip positioned over BDPA near the rim (Fig.~\ref{fig:spec}). 
The Kondo dip close to $E_\mathrm{F}$ is visible (red arrow) and significantly larger than the modulation ('ripples') of the surface-state band caused by the regular herringbone reconstruction of the Au(111) \cite{Chen1998}. 
Typical spectra exhibit a small filled-state resonance around $-0.7$ to $-0.9$~eV and a strong empty-state resonance around $+1.5$~eV marked by black arrows. 
We attribute them to the doubly occupied HOMO and doubly unoccupied LUMO-related states. 

Unfortunately, no distinct SOMO/SUMO resonances are observed in the spectrum. 
This may be due to the steric protection of the SOMO/SUMO, which corresponds to a small tunnelling amplitude into or out of the SUMO/SOMO -- the same reason as for the observed small Fano $q$ factor (see above).
Nevertheless, the positions of SOMO/SUMO can be approximated by a comparison with the spectrum of the pristine Au(111) surface in Fig.~\ref{fig:spec}. 
Close to about $\pm 0.5$~V the conductance of BDPA is reduced below that of the pristine substrate, suggesting two antiresonances (marked by green arrows) that relate to SOMO/SUMO. 
The respective SOMO-SUMO energy gap of $U\approx 1$~eV is consistent with the DFT-calculated value of $U=1.56$~eV obtained for a single BDPA radical in the gas-phase, because a slight narrowing of the gap by a few tenths of electronvolts is commonly observed for organic molecules adsorbed on metal surfaces. 

\subsection*{S3. Kondo screening cloud}

Even the largest characteristic lengths of several nanometers \cite{Manoharan2000,Pruser2011} 
so far obtained from experimental data are much shorter than the theoretical Kondo screening length $\xi_\mathrm{K}= \hbar v_\mathrm{F} / (k_\mathrm{B} T_\mathrm{K})$ predicted to be about 0.1--1~$\mu$m (Fermi velocity $v_\mathrm{F}=\hbar k_\mathrm{F}/ m^* $ and effective mass $m^*$) \cite{Affleck2001,Affleck2008}. 
Recent theoretical studies seem to be just about to resolve this issue, and corroborating that $\xi_\mathrm{K}$ plays no significant role in experimental $\mathrm{d}I/\mathrm{d}V$ datat, but that, rather, the experimentally observed $\mathrm{d}I/\mathrm{d}V$ features decay within a few Fermi wavelengths \cite{Busser2010}.

For the Au(111) surface state ($m^*=0.25\cdot m_\mathrm{0}$, $k_\mathrm{F}=0.162$~\AA$^{-1}$ \cite{Reinert2003}) a value of $\xi_\mathrm{K} = 106$~nm is obtained.
The diameter of our RCEs observed close to $E_\mathrm{F}$ are much smaller than $\xi_\mathrm{K}$.
(For Au bulk electrons with $v_\mathrm{F} \approx 1.4 \cdot 10^{6}$~m/s \cite{Ashcroft1976} one obtains a value of $\xi_\mathrm{K} = 198$~nm).

\subsection*{S4. Spatial decay of the Kondo resonance signal}

\begin{figure}
\includegraphics[width=8.6cm]{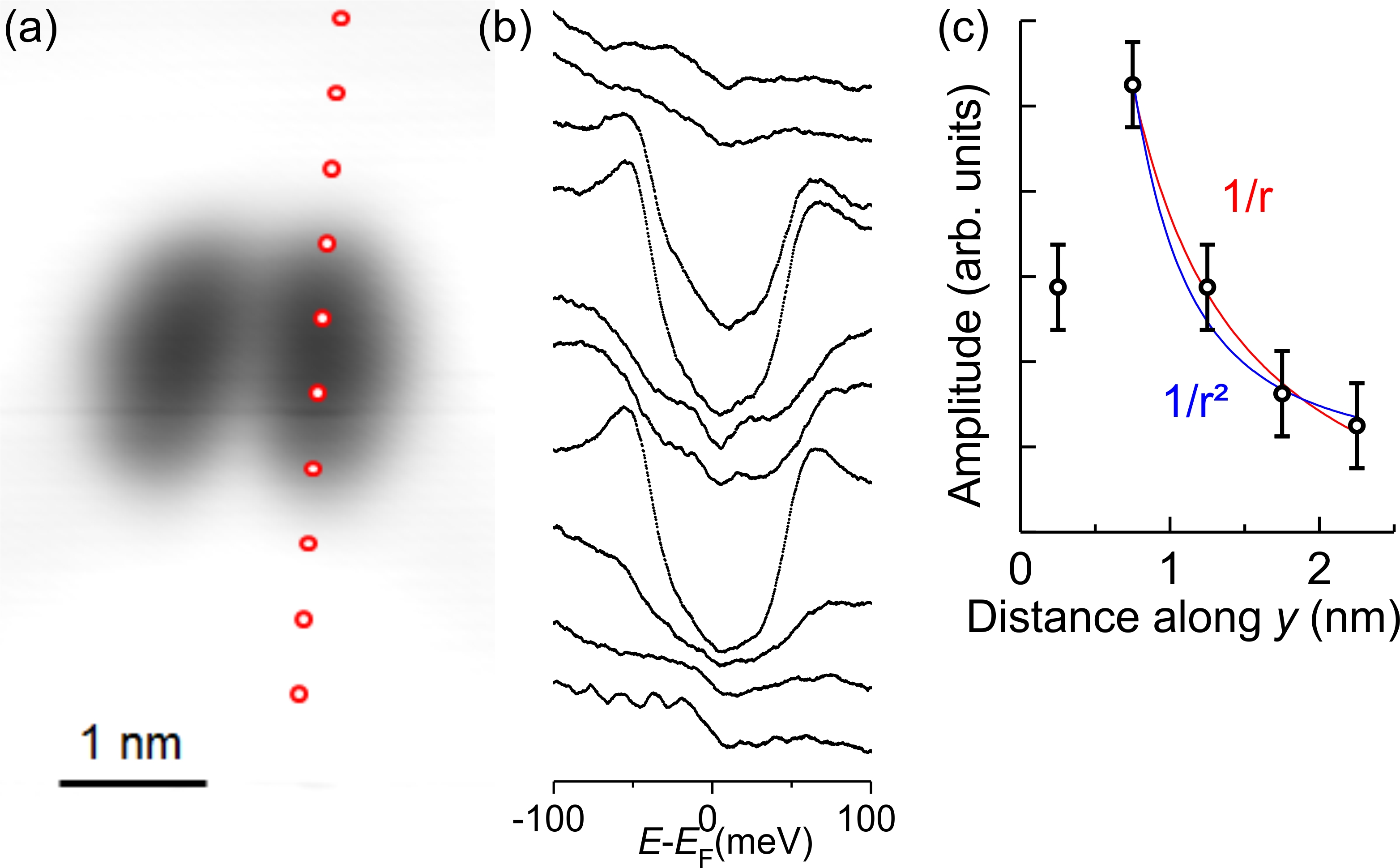}
\caption{\label{fig:decay} \textbf{Distance-dependent decay of the Kondo signal.} (a) STM topo of BDPA dimer. (b) $\mathrm{d}I/\mathrm{d}V$ Kondo spectra recorded at different tip positions as marked by circles. (c) Decay of Kondo amplitude with increasing distance from BDPA; solid lines are fits of $1/r$ and $1/r^2$. }
\end{figure}

Figure~\ref{fig:decay} illustrates the results of distance-dependent STS measurements along a specific high symmetry direction of a BDPA dimer. 
STS spectra were recorded with the STM tip over positions at increasing lateral separation from the BDPA dimer as marked by the circles in Fig.~\ref{fig:decay}a. 
The respective spectra shown in Fig.~\ref{fig:decay}b exhibit the characteristic Kondo dip even at large lateral separations of more than 1.5~nm from BDPA, while the amplitude of the dip  decreases monotonically.
The spatially decaying amplitude values are plotted in Fig.~\ref{fig:decay}c together with the best-fit curve obtained for a decay according to $1/r$ (red curve) with a reduced $\chi^2=1.1$. 
For comparison, we fitted a $1/r^2$ decay as well (blue curve), which results in a significantly poorer agreement with the experimental data ($\chi^2=1.7$). 

\subsection*{References}